# Mesa-top quantum dot single photon emitter arrays: growth, optical characteristics, and the simulated optical response of integrated dielectric nanoantenna-waveguide systems

Jiefei Zhang,[1] Swarnabha Chattaraj,[2] Siyuan Lu[3] and Anupam Madhukar[1,4,a]

[1]*Department of Physics and Astronomy, University of Southern California, Los Angeles, California 90089, USA*

[2]*Ming Hsieh Department of Electrical Engineering, University of Southern California, Los Angeles, California 90089, USA*

[3]*IBM Thomas J. Watson Research Center, Yorktown Heights, New York, 10598, USA*

[4]*Mork Family Department of Chemical Engineering and Materials Science, University of Southern California, Los Angeles, California 90089, USA*

Nanophotonic quantum information processing systems require spatially ordered, spectrally uniform single photon sources (SPSs) integrated on-chip with co-designed light manipulating elements providing emission rate enhancement, emitted photon guidance, and lossless propagation. Towards this goal, we consider systems comprising an SPS array with each SPS coupled to a dielectric building block (DBB) based *multifunctional* light manipulation *unit* (LMU). For the SPS array, we report triggered single photon emission from GaAs(001)/InGaAs single quantum dots (SQDs) grown selectively on top of nanomesas using the approach of substrate-encoded size-reducing epitaxy (SESRE). Systematic temperature and power dependent photoluminescence (PL), PL excitation, time-resolved PL, and emission statistics studies reveal high spectral uniformity and single photon emission at 8 K with $g^{(2)}(0)$ of 0.19±0.03. The SESRE based SPS arrays, following growth of a planarizing overlayer, are readily integrable with LMUs fabricated subsequently using either the 2D photonic crystal approach or, as theoretically examined here, DBB based LMUs. We report the simulated optical response of SPS embedded in DBB based nanoantenna-waveguide structures as the multifunctional LMU. The multiple functions of emission rate enhancement, guiding, and lossless propagation derive from the behavior of the same collective Mie resonance (dominantly magnetic) of the interacting DBB based LMU tuned to the SPS targeted emission wavelength of 980 nm. The simulation utilizes an analytical approach that provides physical insight to the obtained numerical results. Together, the combined experimental and modelling demonstrations open a rich approach to implementing co-designed on-chip integrated SPS-LMU that, in turn, serve as basic elements of integrated nanophotonic information processing systems.

---

[a] Author to whom correspondence should be addressed. Electronic mail: madhukar@usc.edu. Telephone: 1-213-740-4325.

**I. INTRODUCTION**

Realization of on-chip integrated nanophotonic systems comprising light source, light manipulating structures (cavity, waveguide, etc.), and detectors operating down to a single photon level for quantum information processing is a key goal of semiconductor photonics[1,2]. Investigations of vapor phase epitaxially grown semiconductor quantum dots on III-V substrates[3,4] as well as integrated on Si substrates[5] as single photon sources (SPSs) play a central driving role in the progress towards this goal[1,6]. A continuing critical need however is realizing greater control on the uniformity of the spatial ordering of single quantum dots (SQDs) in three dimensions and, even more demanding, their spectral emission. Additionally, the SQD based single photon emitters must be in arrays that are naturally and readily integrable with several co-designed light manipulating elements (LMEs) such as waveguide, resonant cavity, etc., that together we refer here as a multifunctional light manipulating unit (LMU). The resulting interacting SQD - multifunctional LMUs in turn provide the hierarchical building blocks that can be readily interconnected in architectures designed to provide circuit- and system-level functionality. To date, approaches to integrated SQD-LMU structures have been based on either micropillar[7-10] / nanowire[11-14] or 2D photonic crystal[1,15-20] platforms, the former inherently well-suited to architectures exploiting vertical emission of photons while the latter to horizontal emission. Great progress has been achieved using vertical micropillar / nanowire architecture in realizing single photon emission rate enhancement[7-10], purity[8,10,11,13,14], indistinguishability[9,10,14] and photon collection efficiency control[9-14], all important figures of merit for SPSs. For the horizontal architecture, with the advancements in nanoscale lithography and etching over large areas, 2D photonic crystal platform has been shown to manipulate single photon emission (rate, purity etc.) [15-20] and build on-chip interconnected structures[19,20]. In both approaches, so far the reported studies are invariably on pre-selected single QD - cavity / waveguide single units, not array. For the 2D photonic crystal approach, typically the QD-cavity/waveguide unit is built around a single nanoscale semiconductor 3D island-based so-called self-assembled quantum dot (SAQD)[3], chosen from an inhomogeneous and spatially random distribution of SAQDs formed during lattice-mismatch driven strained epitaxy on a semiconductor substrate[3,6] (most commonly InGaAs on GaAs(001) substrate). Thus the limiting factors now are the deterministic positioning and spectral matching of quantum dot single photon source to the 2D photonic crystal platform-based optical modes invoked for coupling to the SQD transition[1]. Quantum dots with predetermined formation position and spectral uniformity that falls within the tuning range of on-chip scalable spectral tuning technologies[19,21], is an essential requirement.

To facilitate movement towards implementation of the needed on-chip integrated nanophotonic systems, we report here on systems comprising spatially-ordered SQD based SPS array with each SPS surrounded by a dielectric building block (DBB) based *multifunctional* light manipulation *unit* (LMU)[22]. For the SPS array, we report on the fabrication and optical characterization of spatially-ordered and spectrally-uniform SPS 5x8 array (Fig.1a) based on GaAs(001)/InGaAs/GaAs SQD residing on nanomesa tops synthesized through the approach of substrate-encoded size-reducing epitaxy (SESRE)[23,24,25] that naturally lends itself to on-chip integration with light manipulation units. The array exhibits an average $g^{(2)}(0)$ of 0.19±0.03 at 8 K reflecting good quantum confinement. The emission at 935.3±8.3 nm reveals uniformity significantly improved over the commonly employed SAQDs (also known as Stranski-Krastanov QDs) in the on-chip integrated 2D photonic crystal based systems reported to date[1,17-19].

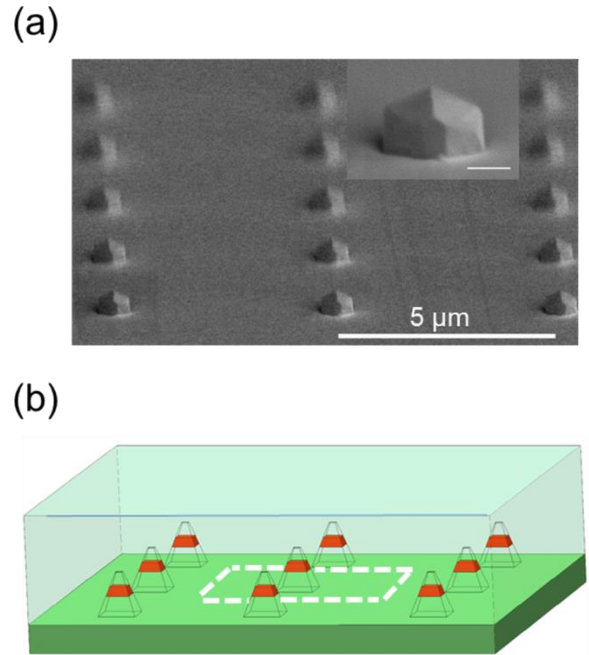

Fig.1. (a) A 60° tilted SEM image of the SESRE-grown InGaAs MTSQD array with the MTSQD residing on the top of each nanomesa. The inset is a magnified image of a MTSQD-bearing single nanomesa (scale bar of 300 nm) (b) Schematic showing the planarization of MTSQD arrays with continued overgrowth (i.e. GaAs) which enables the post-processing and fabrication of LMUs around each MTSQD (marked within the area of white dot line) acting as a functional unit cell on chip.

Concerning the critical need for on-chip integration with light manipulating elements, we note that in the SESRE approach, the SQD-bearing nanomesa arrays are readily planarized by overgrowth of an appropriate layer (Fig.1b), GaAs for the case of GaAs/InGaAs SQD array reported on here. The resulting system with known locations of buried SPSs in an array provides the natural starting point for the co-designed nanolithographic fabrication of the desired light manipulating structures in the overlayer. For the current Bragg diffraction based 2D photonic crystal approach it will overcome the limitation imposed by the random location of SAQDs, faced so far in realizing interacting array configurations essential for creating platforms for designed SQD-SQD couplings[1].

Equally of value, a system of buried spatially ordered array of SPSs also opens opportunities for experimental investigations of the recent theoretical explorations of the potential of subwavelength DBB based structures for providing light manipulation functions based upon Mie resonances as another approach for horizontal light emission and propagation architectures, with much smaller overall footprint[26-30] compared to 2D photonic crystal approach. Motivated by such holistic perspective, we have undertaken theoretical examination of the optical response of integrated architectures in which each emitter of a regular array is coupled to a DBB based composite LMU that *simultaneously* provides nanoantenna and waveguiding functionalities, as symbolically captured in Fig. 2 by an individual row comprising the pyramidal structure (denoting the single quantum dot) and the blue dielectric building blocks.

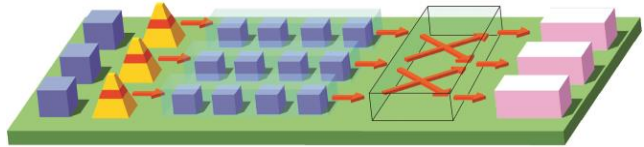

Fig.2. Schematic showing an array of SESRE-grown mesa-top quantum dots (pyramids), on-chip integrated with lossless dielectric building block (DBB) based LMU (blue elements) that provide multiple functions (resonant cavity/nanoantenna/waveguide) and can be integrated with beam splitters and modulators as well as detectors (purple elements) to create photonic information processing systems[22].

Accordingly, we report here the simulated optical response of GaAs based nanoantenna-waveguide composite structure coupled to a dipole emitter at 980 nm. It shows an enhancement of the emission rate (i.e. the Purcell factor) by 7 and lossless propagation along the structure. These multiple functions are derived simultaneously from the collective Mie resonance[29,30] of the interacting GaAs DBBs comprising the LMU (nanoantenna-waveguide integral structure). The simulation approach developed employs an analytical description of light scattering whose numerical implementation is highly efficient to enable co-design of the integrated SPS- multifunctional LMUs. We note that integrated structures such as an individual row comprising the SQD and LMU in Fig. 2 serve as the basic units which, appropriately interconnected, will enable the next level of hierarchical design enabling SQD-SQD communication, and thus the creation of optical circuits (Fig.2) for applications needing many photons down to single. The combined experimental findings on the SESRE single quantum dot array as SPS and modelling of the optical response of DBB based multi-functional LMUs integrated on-chip with such SPSs reported here open a new and rich approach to exploring and fabricating minimal footprint on-chip integrated SPS-light manipulating multi-functional structures.

Substrate-encoded size-reducing epitaxy[23] is an approach to growth-controlled spatially-selective formation of nanostructures that exploits growth on designed *non-planar* patterned substrates, i.e. patterned structurally such that the tailored surface curvature induces surface stress gradients (capillarity) that direct adatoms during deposition preferentially to mesa tops[23-24, 31-35] or valleys[23,36] / recesses[37,38] for selective incorporation through control on the relative kinetics of adatom incorporation on the contiguous facets present in the designed curvature[23, 31-35]. For pattern designs that induce net migration from the side facets to the mesa top, preferential incorporation at the mesa-top leads to growth-controlled mesa size reduction, enabling *in-situ* preparation of contamination and defect-free nanomesa arrays from the as-patterned array via homoepitaxy. Hence substrate-encoded size-reducing epitaxy[23]. For etched recesses, net migration into the recess expands the initial bottom area under conformal growth[23, 37-38]. Subsequent heteroepitaxy, lattice matched or mismatched, then enables synthesis of quantum confined nanostructures such as quantum wires and dots on pristine nanotemplates[23]. For the technologically important (001) surface oriented substrates of the tetrahedrally-bonded semiconductors of groups IV, III-V, and II-VI, the <100> edge orientations of square mesas provide four-fold symmetry and thus potentially symmetric migration of adatoms from the sidewalls to the top (Fig.3a) to reduce, in-situ, the as-patterned mesa top size to the desired size utilizing homoepitaxy under controlled growth kinetics[23,24,31-33]. Indeed, depending upon the chosen as-etched sidewall crystallographic planes, the size-reducing growth can offer more than one pinch-off stages[33], thereby allowing control on not only size but also the shape of the QD formed subsequently via heteroepitaxy as depicted in Fig.3 (b) and (c). Illustrative examples of mesa-top QDs with {101} sidewall facets, taken from ref. 39, are shown in Fig.3(d) for the lattice-matched system GaAs(001)/AlGaAs[23,31] and in Fig.3(e) for the highly lattice-mismatched (7%) system GaAs(001)/InAs[32,33], both synthesized using solid-source molecular beam epitaxy (SSMBE). The flat morphology and the absence of defects for lattice mismatched QD[32] are due to a substantial strain relaxation on a sufficiently small nanoscale mesa owing to the presence of mesa free sidewalls as demonstrated in multi-million atom molecular dynamics simulations[34,35]. It is important to note that growth on mesatop, unlike in recess, enables accommodation of large lattice mismatch strain without generation of defects owing to the mesa free surfaces during growth and is thus suitable for a wide variety of material combinations covering optical emission from the UV (such as in the III-Nitrides) to the midinfrared (such as in the small bandgap InSb). The SESRE quantum dots reported here exploit the shallow {103} sidewalls[24]. Finally, we note that SESRE not only permits control on mesa-top SQD shape and size, it also allows controlled vertical stacking and coupling of QDs, as desired[31,32]. Moreover, unlike other template-directed growth approaches that often suffer from formation of unintended parasitic nanoscale structures, SESRE carried out on properly chosen mesa orientations and shapes under careful control of growth kinetics leads only to the formation of predictable and intended structures[23,32,33]. The controlled formation of uniform SESRE QD array, however, requires atomistic level control on the growth kinetics and nanometer precision of the starting mesa lateral size. With the development of nanolithographic technologies that have enabled nanometer precision control over large areas, the SESRE approach offers a path worthy of exploration as demonstrated below. Henceforth we denote these surface-curvature stress-gradient-driven QDs as mesa top single quantum dots (MTSQDs).

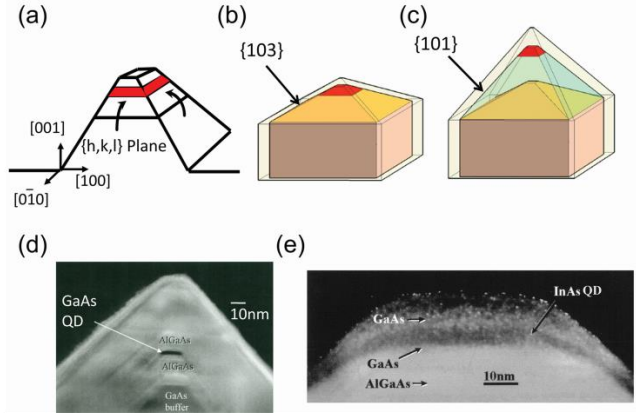

Fig.3. (a) Schematic of a (001) top square mesa with <100> edge orientations and {h,k,l} sidewalls. The arrows indicate atom migration from the sidewalls to the top leading to size-reducing epitaxy. The red box depicts the flat-top pyramidal quantum dot formed at the designed stage before mesa pinch-off by {103} planes (panel b) or later by {110} planes (panel c) to enable flat-top pyramidal quantum dot (red) formation. Panels (d) and (e), taken from ref. 39, show respectively cross-sectional TEM images of a SESRE grown lattice matched AlGaAs/GaAs/AlGaAs quantum dot with base 11 nm and height of 2 nm and a 7% lattice mismatched GaAs/InAs/GaAs quantum dot of base ~40 nm and height ~3 nm at the apex of a nanomesa bounded by {101} planes.

To date, a large part of the single photon studies[1,6,40-42] have employed lattice-mismatch strain driven spontaneously formed defect-free 3D island QDs dubbed SAQDs[3,4]. The work horse has been the GaAs(001)/InGaAs material system that, on planar GaAs(001) spontaneously forms 3D islands for ~1.6 ML deposition of InAs. The SAQDs have also been incorporated in coupled SPS-cavity[7-10,17,19,20], SPS-waveguide[15,16,18], and SPS-cavity-waveguide[19] systems. However, the random nature of SAQD formation in space and time, inherent to their spontaneous initiation, leads to randomness in their location and considerable variation in island size, shape, and composition that results in unacceptable spectral non-uniformity in the SAQD emission[43-47]. Their random location has prevented integration with optical elements such as cavities, waveguides, etc. in ordered arrays as required for photonic systems. The large spectral non-uniformity makes it difficult to spectrally match QD with the light manipulating elements within the tuning range of the well-established spectral tuning technique[19,21]. Another class of SPS is based on QDs in nanowire[12-14,48-50] structures typically grown via the metal nanoparticle seeded vapor-liquid-solid (VLS) growth mechanism. The nanowire diameter is controlled by the seed particle diameter and in turn the QD lateral size is controlled by the nanowire diameter[48,49]. Thus for sufficiently small diameters, even for highly lattice mismatched combinations such as GaAs/InAs or InP/InAsP the shape of the QD is not island like (i.e. not SAQD like) but rather flat as in the SESRE quantum dots since sufficiently small diameter allows these nanowire embedded QDs to also be strain relaxed without defects, the same advantage as the SESRE grown QDs (Fig.3e). However, unlike the SESRE approach, the spatially random distribution of the seed nanoparticles leads to random location of the QD bearing nanowires.

To achieve spatial regularity, a popular approach is deposition on substrates with appropriate masks - typically oxides – with patterned etched holes, leaving the underlying semiconductor substrate *planar*[23]. The growth is typically done using all gaseous molecular sources such as in gas source MBE or metal organic chemical vapor deposition (MOCVD) as the slow dissociative reaction kinetics and fast desorption kinetics of the relevant molecular species on the oxide enables selective growth in the holes. The lateral size and depth of the holes has been exploited for selective placement of quantum dots, including single[51-53]. However, such growth, being spontaneous, is sensitive to the fluctuations in the size and shape of the holes and does not allow the needed control on the shape and size of QDs formed. Most recently the holes-in-oxide-mask approach has been combined with the metal nanoparticle-seeded VLS growth[48-49] by creating the seed metal nanoparticles selectively in the holes. The *non*-SAQD quantum dots in such nanowires have been demonstrated to be single photon sources[48,50]. Moreover, the nanowire bearing the QD can also be designed to boost the collection efficiency of the emitted photons from the QD [12,13,50]. Nevertheless, certain challenges remain to be tackled, including the incorporation of metal point defects that adversely impact the optical properties of the QDs[54], the fluctuation of nucleation events underlying VLS growth, and variation in the seed nanoparticle sizes[55] which together make predictable control on QD vertical location and emission wavelength difficult. Given the current situation, with the maturing of nanolithography enabling precise control on starting nanomesa shape and size, here we revisit the SESRE approach and report the realization of MTSQD SPS array that is efficient at the elevated temperature of 77 K and whose spectral emission is significantly more uniform than typical SAQDs and nanocrystal quantum dots and comparable to the uniformity of the most recently reported nanowire-embedded QD system[56]. The SESRE array is, however, more naturally compatible with the subsequent monolithic integration with light manipulating structures as mentioned previously and discussed in the following.

To manipulate *on-chip* the light emitted by the SPS, starting with enhancing the emission rate (the Purcell effect), spatially directing the emitted photons by modifying the local photon density of states, and propagating the photons without loss to the desired destination, the combination of SAQD and 2D photonic crystal platform has been exploited [15-20]. Given the nature of the materials comprising the self-assembled island semiconductor quantum dots, the use of the 2D photonic crystal platform is a natural choice. The physical phenomenon underpinning photonic crystals is Bragg diffraction in perfect periodic bulk-like structures and the ability to derive desired functionality (such as high Q resonant cavity and waveguiding) by introducing departures from perfect crystallinity through controlled introduction of particular types of defects in the photonic lattice[1,19]. By contrast, Mie scattering of optical photons by a suitable collection of interacting subwavelength size DBB usually leads to comparatively low Q factors/ lower degree of confinement of electromagnetic modes, but can provide functionalities such as nanoantenna and waveguiding with a much smaller on-chip foot-print and potentially greater flexibility in implementation. For example, propagation of light on subdiffraction scale exploiting collective Mie resonance in linear array of DBBs has been demonstrated resulting in lossless waveguide structures with much smaller footprint compared to the conventional photonic crystal based systems[26,57,58]. Additionally, using the symmetries of the Mie resonant modes, DBB based nanoantenna structures[27] have been shown to enhance the emission rate (Purcell effect) of a dipole like source. Recently we have proposed the use of the collective properties of the different multipole modes of high index DBBs (such as GaAs, Si, TiO$_2$) in QD-DBB integrated systems (Fig.2) to simultaneously implement multiple functions[29,30] such as focusing excitation light, enhancing emission rate, guiding and propagating the emitted photons. Simultaneous implementation of such broad range of functions on micron scale footprint carries high payoff towards the development of integrated quantum nanophotonic system design. Guided by this potential, we report here simulations of the Mie scattering based optical response of GaAs DBB based light manipulating structures that are natural for the class of GaAs/InGaAs/GaAs MTSQD arrays that we present here and are developing further.

## II. Results

### A. Mesa Top Single Quantum Dot Array Synthesis via SESRE

For the synthesis of the MTSQD arrays, electron beam lithography and wet etching are combined to pattern GaAs(001) substrates with 5x8 arrays of nanomesas with lateral sizes between ~100 nm and ~500 nm, height ~500 nm, and vertical side walls. Alongside the patterned nanomesa region, an L-shape unpatterned region is left on the substrate to serve as a reference for calibrating and controlling gallium and indium fluxes (and thus growth rates, composition, and thickness) to atomic layer deposition precision using real-time RHEED (reflection high energy electron diffraction) pattern and intensity dynamics in our MBE system. The starting mesa height is chosen to be sufficient for the growth of a GaAs buffer layer that not only buries the residual contamination and defects left after etching and native oxide desorption but also provides adequate mesa top size reduction control prior to the stage of the quantum dot deposition on top. The chosen as-patterned mesa size range allows us to investigate different stages of pinch-off in the same growth run since at the same stage of deposition mesa pinch-off occurs later on mesas with larger starting lateral size. Details of the substrate patterning, growth conditions, and grown structure can be found in ref. 24. The single photon emission results reported here are from a 5x8 array of flat-top pyramidal MTSQDs with {103} side planes (Fig.3(b)) formed with 4.25 ML In$_{0.5}$Ga$_{0.5}$As deposited on size-reduced GaAs(001) mesa tops with as-etched nanomesa size of ~430 nm. The inset in Fig.1(a) is a magnified image of a typical post-growth mesa structure containing a single flat-top InGaAs quantum dot bounded by {103} planes at the stage of encapsulation by GaAs as illustrated in the schematic shown in Fig. 3(b). These MTSQDs have base length of ~13 nm and height ~3 nm as estimated from growth evolution. Note that, unlike SAQDs, there is no wetting layer involved in the formation of these MTSQDs. Thus the confinement of the electronic states is expected to be stronger, as borne out by the optical measurements reported below. The inter-mesa separation is chosen to be 5 μm to facilitate optical characterization of each MTSQD in the array using our micro-photoluminescence setup as discussed below.

**B. Optical Response and Spectral Uniformity of MTSQDs**

The photoluminescence (PL) from every MTSQD in the 5x8 array were measured at 77.4 K using our home-built micro-photoluminescence (μ-PL) system. Excitation light from an optical fiber coupled 640 nm 80 MHz diode laser (<100ps pulse width, PicoQuant model LDH-P-C-640B), focused down to ~ 1μm diameter through a 40× NA 0.65 objective lens, is vertically incident on the sample mounted in a Janis ST-500 continuous flow microscope cryostat. The PL emission from a single MTSQD, collected through the same microscope objective, is filtered by a set of dichromatic and long pass filters and then a 30 cm focal length spectrometer (Acton Spectro300i) and detected by silicon avalanche photodiodes (APDs, PicoQuant model τ-SPAD). The band pass of the spectrometer is set at 0.4 nm (~570 μeV) for photoluminescence measurements at 77.4 K. The μ-PL spectra from all 40 MTSQDs in the 5x8 array were recorded. The spectrum from a representative MTSQD in the array (Fig. 4(a)) shows a peak at 929 nm with full width at half-maximum (FWHM) of 1.3 nm (1.85 meV) limited by thermal broadening at 77K and Stark shift induced by possible charge fluctuations at the surface of the capping GaAs layer[59]. The excitation power ($P$) dependence of the PL intensity ($I$) shown in Fig. 4(b) reveals $I \sim P^{0.91}$, a nearly linear increase indicative of single exciton emission. The PL from these MTSQDs saturates at an order of magnitude lower saturation power (~30 nW and 2.4 W/cm$^2$) compared to that reported for other spatially-ordered QDs such as InGaAs/AlGaAs QDs in valleys (holes) of structurally patterned GaAs(111)B substrate[60], InAsP/InP nanowire QD[48] and GaN/AlGaN nanowire QD[53]. This suggests that these MTSQDs synthesized through the SESRE approach may have high carrier capture rate. We note that the SESRE approach enables growth-controlled formation of MTSQDs without any unintentional structure to compete with the QD for carrier capture unlike the unintended parasitic heterostructures formed in nanowire QD structures[55]. Additionally, time-resolved PL measurements, shown in Fig. 4(c), reveal a mono-exponential intensity decay with a life time of 0.8ns comparable to that of typical InGaAs SAQDs[4,61]. The measured PL intensity at 77.4 K at the low (nW) powers employed and the short (ns) PL decay lifetime of MTSQD single excitonic transition indicate good quantum confinement of excitons.

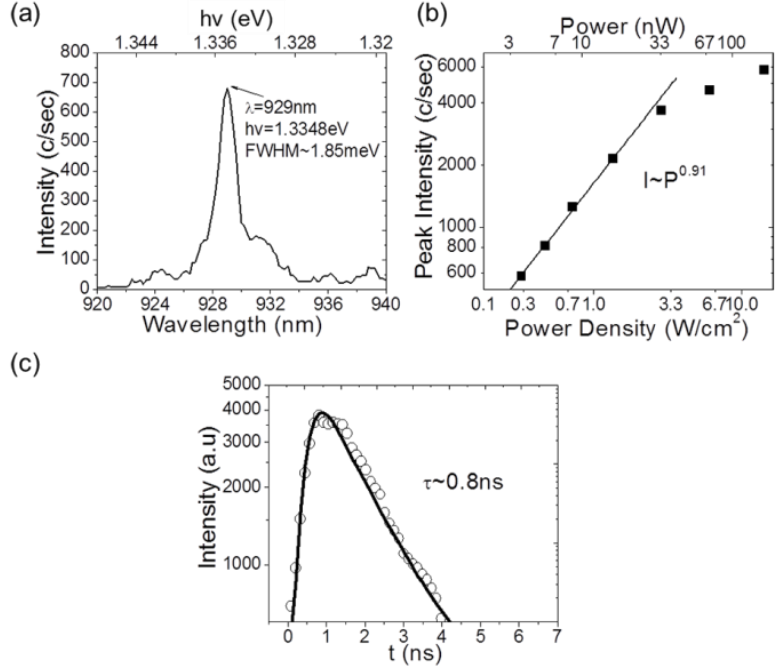

Fig. 4. Optical behavior of MTSQD.(a) Photoluminescence (PL) spectra of a MTSQD (3,5) (labeled by its row and column number in the array) collected at 77.4 K and excitation power ~4 nW (~0.32 W/cm$^2$). (b) Power dependence of this MTSQD's PL peak intensity at 77.4 K. A fit (black line) to the data shows $I\sim P^{0.91}$. (c) Time-resolved PL (TRPL) data of this single MTSQD collected at the same condition as in (a). A fit to the data (black line) reveals single exponential decay with a lifetime of ~0.8 ns.

The temperature dependence of the PL was also investigated and the measured shift of the PL intensity peak of MTSQD (3,5) with temperature is shown in Fig. 5(a). It is found to follow the calculated temperature dependence of the In$_{0.5}$Ga$_{0.5}$As band gap change (solid line, Fig. 5(a)) obtained from the following expression for In$_x$Ga$_{1-x}$As alloy band gap[62],

$$\Delta E(x,T) = \Delta E^{InAs}(T)x + \Delta E^{GaAs}(T)(1-x) \quad (1)$$

in which $\Delta E^{InAs}(T)$ and $\Delta E^{GaAs}(T)$ are of the Varshni form[63],

$$\Delta E(T) = \frac{\alpha T^2}{T+\beta}, \quad (2)$$

with input parameter x= 0.5, $\alpha^{InAs}$= -4.19x10$^{-4}$ eVK$^{-2}$, $\beta^{InAs}$= 271 K and $\alpha^{GaAs}$= -5.8x10$^{-4}$ eVK$^{-2}$, $\beta^{GaAs}$= 300 K taken from Ref. 62 corresponding to the bulk (strain-free) InAs and GaAs, respectively. Furthermore, the integrated PL intensity ($I$) plotted in Fig. 5(b) in the range 77.4 K to 150 K shows an exponential dependence on the inverse of temperature, $I\sim\exp(E_{act}/k_BT)$, with $E_{act}$ of 40±2 meV, revealing an activated electron escape through the first excited state contributing to the non-radiative decay. These findings not only further support the origin of the emission as being from the InGaAs QDs but also reveal the good spatial confinement of electron and hole wavefunctions in these MTSQDs.

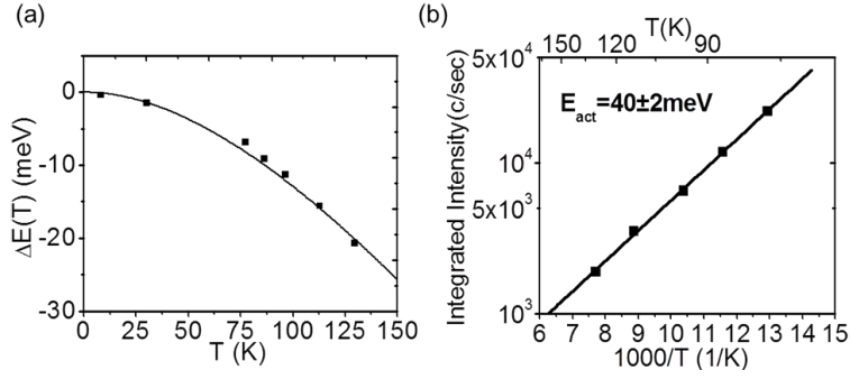

Fig. 5. Temperature dependent PL of MTSQDs. (a) MTSQD (3,5) PL peak energy shift with temperature. The shift follows closely the temperature dependence of the $In_{0.5}Ga_{0.5}As$ bandgap change (black line). (b) Temperature dependence of this MTSQD's integrated PL intensity from 77.4 K to 150 K. The fit (black line) to the data reveals an exponential dependence on inverse temperature with an activation energy of 40±2 meV representing carrier escape from the QD.

To shed further light on the potential nature of the states involved in the PL, we examined the PL excitation (PLE) behavior. The PL intensity for MTSQD (3,5), measured with the detection energy set to its PL peak at 929 nm and an acceptance window of 0.2 nm, as a function of the energy difference between excitation and detected photon energy is shown in Fig. 6. Of the four prominent peaks observed (labeled as $P_1$-$P_4$), position of peaks $P_2$ (25.8 meV) and $P_3$ (33.3 meV) are independent of the detection energy and thus are consistent with the expected behavior of optical phonon-assisted emission. The peaks $P_1$ and $P_4$ shift their position with the detection energy and are consistent with the presence and involvement of excited hole ($P_1$) and electron ($P_4$) states. Further PLE studies with improved spectral resolution and as a function of temperature lowered down to LHe will provide added information which, in conjunction with calculated electronic structure, will enable the identification of the states predominantly involved.

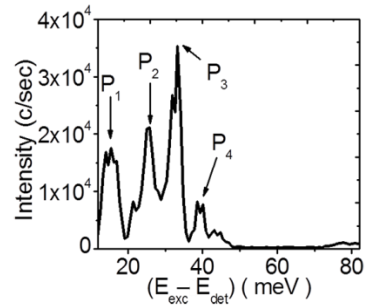

Fig 6. Photoluminescence excitation (PLE) spectrum of MTSQD (3,5) with detection at QD emission peak 929nm with acceptance window of 0.2 nm. Detected PL intensity- displayed as a function of the difference of excitation energy and detected photon energy- shows four peaks at 15.5 meV, 25.8 meV, 33.3 meV and 39.4 meV marked, respectively, as $P_1$, $P_2$, $P_3$ and $P_4$.

To address the spectral uniformity of emission from these MTSQDs, as noted above, PL spectra were collected from *all* 40 MTSQDs in the array. Only two out of the 40 are non-emitting (marked as black boxes in Fig. 7) and the intensity from the rest 38 MTSQDs varies within a factor of 4 over the whole array, as shown color coded in Fig. 7(a).

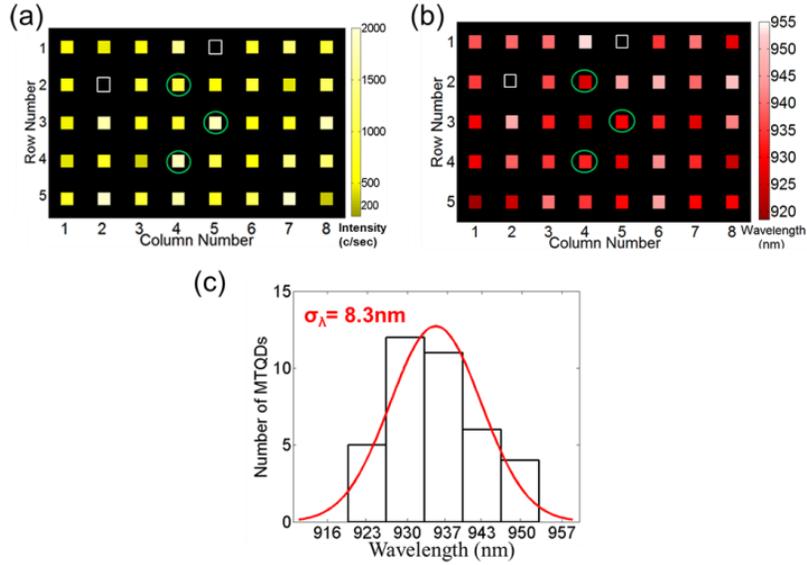

Fig.7. Spectral uniformity of an MTSQD array.(a) PL peak intensity and (b)PL peak wavelength of each MTSQD in the 5x8 array shown as color coded blocks. The two non-emitting MTSQDs are marked as black blocks with white outlines. MTSQDs marked by green circles are those examined for $g^{(2)}(\tau)$. (c) Histogram of PL peak wavelengths from the 38 emitting MTSQDs in the 5x8 array covering an area of 1000 μm². The standard deviation of the wavelength is 8.3 nm. The red line represents the Gaussian fit to the distribution.

In addition to their intensities, the PL peak wavelengths of the 38 MTSQDs in the array are shown color-coded in Fig. 7(b) and plotted as a histogram in Fig. 7(c). These MTSQDs show average PL peak position at 935.3 nm with a standard deviation, $\sigma_\lambda$, of 8.3 nm which is <1% of the emission wavelength. This is a significant improvement over typical SAQDs[3,4] and NCQDs (nanocrystal QDs)[64]. The inherently built-in growth control on QD shape (flat-top pyramidal shape with controlled crystallographic side planes) enables the high spectral uniformity of MTSQDs. The MTSQDs with emission at the shortest (919 nm) and longest (951 nm) wavelengths lie on the outer row and column in the 5x8 array. The observed variation of the MTSQD emission wavelengths most likely originates from the variation of the as-etched mesa size. Such mesa size variation causes variation in the starting QD base length (at the end of the homoepitaxy prior to InGaAs deposition) and thus variation in QD height and indium composition. Indeed, concerning the latter, binary InAs MTSQDs have revealed a significantly narrower spectral emission[24]. No attempts were made to optimize these for the MTSQDs in these first studies. Through improved control on the starting as-etched mesa sizes and control on gallium and indium migration lengths by optimizing growth kinetics, spectral uniformity of the MTSQDs can be further enhanced. Efforts in these directions are underway. Overall, MTSQDs synthesized using SESRE approach provide not only spatially-ordered QDs but also considerable control on spectral uniformity through control on QD shape, size, and composition, making them suitable for applications that require known QD locations and optical response.

**C. Single Photon Emission**

The similar confinement potential amongst these ordered MTSQDs as manifested in their PL behavior and spectral uniformity indicates that such MTSQDs array should have similar single photon emission behavior. To explore these MTSQDs as SPSs, the second order correlation function, $g^{(2)}(\tau)$, of the emitted photons was obtained from measured coincident photon events using a Hanbury-Brown and Twiss setup. All data were taken with 640 nm above gap excitation with 90 ps pulse at a repetition rate of 80 MHz at lowest possible power and detected at each MTSQD's PL peak wavelength with the spectrometer bandpass set at 0.4 nm. The photoluminescence from the MTSQD is spectrally filtered and directed through a 50/50 beam splitter to two Si APD detectors with photon timing resolution of 350 ps, one at each end of the two beams. The timing of the pulses from the two detectors is registered using two constant fraction differential discriminators and fed to a time-to-amplitude (TAC) convertor as the start and stop signal (stop signal delayed by a delayer). The TAC outputs are read by a multi-channel analyzer to generate a histogram of coincidence photon detection events.

The behavior of $g^{(2)}(\tau)$ at liquid nitrogen temperature has been examined for several MTSQDs but the prohibitive expense of liquid helium has restricted study of their single photon emission behavior at lower temperature (~8 K) to only three MTSQDs (green circles, Fig. 7(a) and (b)) in the 5x8 array. Figure 8 shows the histograms of as-measured (raw data) coincident detection of photon emission at 77 K with excitation power ~4 nW (panels (a) and (b)) and at 8K with excitation power ~8 nW (panels (c) and (d)) for two of the three MTSQDs (labeled by their row and column number in the array). The 77 K *as-measured* data for MTSQD (3,5) and (2,4) give normalized $g^{(2)}(0)$ of 0.43 and 0.38 respectively, determined by the ratio of τ=0 peak area to the average of the nonzero peak areas from such histograms (Fig. 8(a) and (b)). After detector dark count subtraction the extracted $g^{(2)}(0)$ values, shown in parenthesis, are 0.38 and 0.32, respectively. The 8 K $g^{(2)}(\tau)$ behavior of these two MTSQDs are shown in Fig.8 (c) and (d), respectively. The corresponding $g^{(2)}(0)$ extracted from as-measured and after detector dark count subtraction data are 0.20 / 0.15 and 0.25 / 0.19. The three MTSQDs examined at 8K give on average $g^{(2)}(0)$ ~ 0.19±0.03 extracted from data with the detector dark count subtracted. Thus these SESRE based MTSQDs provide single photon emission comparable to that of InGaAs SAQDs reported in the literature up to 77 K[65,66] but with control on the QD position and significantly improved spectral uniformity. These attributes, and the ease of growing a planarizing overlayer, make SESRE MTSQDs well suited for nanophotonic on-chip integrable single photon source (Fig.1b).

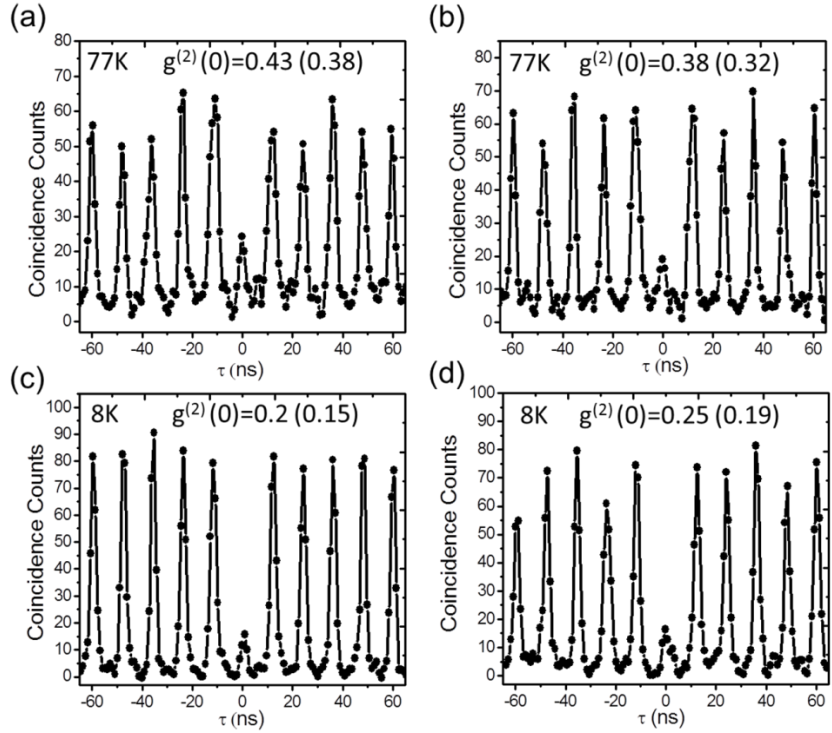

Fig. 8. Single Photon emission from MTSQD array. The 77 K (panels a and b) and 8 K (panels c and d) *as-measured* coincidence count histogram of MTSQD (3,5) and MTSQD (2,4), respectively. The intensity autocorrelation $g^{(2)}(0)$ values shown are extracted from the as-measured data and the values in parenthesis are extracted from data after detector dark count subtraction.

From the lowest excitation power (~4 nW or ~0.32 W/cm$^2$) $g^{(2)}(\tau)$ measurements an exciton lifetime of 1.1±0.3 ns is extracted and is consistent with the MTSQD lifetime obtained from the time-resolved PL measurements illustrated in Fig. 4(c). The short (~1 ns) excitonic lifetime in these MTSQDs can be further shortened by their integration with a cavity or nanoantenna structure to potentially allow triggered single photon emission from regular arrays at greater than GHz operating frequencies, important for high-speed information processing applications.

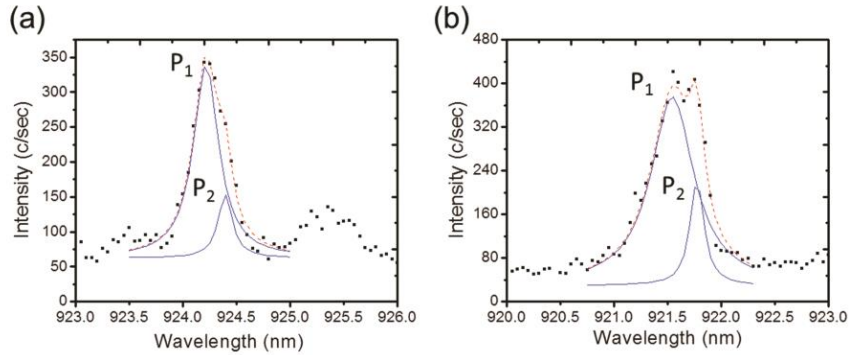

Fig.9. Photoluminescence (PL) spectra of (a) MTSQD (3,5) and (b) MTSQD (2,4) collected at 8 K with 0.25 nm spectral resolution and ~5 nW excitation power. Black dots represent the raw data and the red dashed line shows a fit with two individual Lorentzian peaks ($P_1$ and $P_2$) shown as the solid blue curves.

At low excitation powers (<10 nW), the nonzero g$^{(2)}$(0) value is indicative of more than one initial state of the electron participating in the creation of the emitted photons. Indeed, the PL behavior of all investigated MTSQDs at 8 K, shown here for MTSQDs (3,5) and (2,4) in Fig.9(a) and 9(b) respectively, reveals a lineshape that requires fitting of at least two Lorentizian peaks separated by ~ 0.25 nm (363 μeV) and a peak area ratio ~5. Independently, theoretical analysis of the g$^{(2)}$(0) for two QD states emitting with intensity $I_1$ and $I_2$ and both within the detection window gives,

$$g^{(2)}(0) = \frac{2<I_1><I_2>}{(<I_1>+<I_2>)^2}. \tag{3}$$

Using the intensity ratio of ~5 for the two peaks extracted from the data of Fig.9, a g$^{(2)}$(0) of ~ 0.3 is obtained from Eq.(3), consistent with the measured g$^{(2)}$(0) values. We note that the separation of the two peaks is consistent with the range of fine structure splitting of bright excitons reported for the InGaAs/GaAs SAQDs of sizes and shape similar to these MTSQDs[58,67]. Further investigations at low temperature, including the polarization dependence of the PL, are planned to ascertain the physical origin of the two peaks.

**III. Integration of MTSQD with dielectric nanoantenna and waveguide**

The controlled spatial location along with the spectral uniformity of the MTSQD SPS array makes it ideally suited for integration with passive on-chip optical elements (Figure 2) to enhance excitation and emission rate of individual MTSQDs as well as to guide and propagate the emitted single photons aimed at realizing on-chip optical quantum information processing. Experimentally this is enabled by overgrowth of the appropriate material, here GaAs, on the MTSQD array until planarization. The resulting system of buried SQDs with known spatial geometry is ideally suited for subsequent lithographic fabrication of optical elements using either of the two current approaches: micropillar that enables waveguiding and collection in the vertical geometry[7-10], and particularly the 2D photonic crystal platform that enables manipulation and guidance of the single photons in the horizontal geometry[15-20,68]. Here we investigate a third approach involving the use of the collective resonance of designed systems of interacting nanoscale DBBs that allow manipulation of light on the micron scale beyond the usual diffraction limit[57]. In fact, simultaneous implementation of the needed multiple functions of enhancing the emitted photon rate, guiding, and propagating it without loss of energy or information to the desired location is achieved by exploiting the collective multipole Mie resonances of the high index DBB based structures co-designed with the MTSQD[29,30]. This is unlike realizing similar functions in conventional 2D photonic crystal structures that require larger on-chip footprint to enable Bragg diffraction. Moreover, this approach provides flexibility in choosing the materials of the DBBs that are compatible with the lithographic integration process, e.g. GaAs DBBs can be used for monolithic integration with the SESRE grown GaAs/InGaAs/GaAs MTSQD array.

**Simulation of the optical response of dielectric multifunctional light manipulation units**

To guide the fabrication of such on-chip integrated nanophotonic systems, we undertook simulations of the collective Mie resonance based multifunctional optical response of a nanoantenna-waveguide unit made of GaAs DBBs (Fig.10(a)). We demonstrate here simultaneous MTSQD emission rate enhancement, local nanoantenna effect, and lossless waveguiding of the emitted photons. The emission wavelength of the MTSQD is chosen at 980 nm, same as targeted in the experimental efforts guided, in part, by the availability of silicon avalanche photodetectors that provide sufficient sensitivity. The GaAs DBBs (refractive index 3.5) are approximated as spheres since spherical symmetry allows analytical solution for the Mie scattering which is helpful for gaining physical insight and thus complements finite-difference time-domain (FDTD) method based numerical simulations capable of finding solutions for DBBs of shapes with lower symmetry. The LMU consists of an array of 40 GaAs DBBs (blue spheres, Fig. 10(a)) with radius 129 nm and center-center spacing of 300.6 nm and the purple sphere of radius 131.7nm with an 80 nm surface-to-surface gap to the nearest blue DBB. The radius 129 nm for GaAs DBB corresponds to the magnetic dipole mode ($TE_{1,\pm1}$) of the individual DBBs- exploited here for the nanoatenna and guiding effects- resulting in the collective resonance of the interacting array to be at 980 nm, the emission wavelength of the MTSQD. An electrical point dipole approximating the MTSQD transition is placed 40 nm away from the surface of the purple DBB as well as from the blue DBB. The transition dipole is chosen to be transverse (either along the y or z directions in Fig.10) to the axis of the nanoantenna-waveguide DBB array to maximize its coupling to the transverse magnetic dipole modes of the DBBs exploited to implement the light manipulating functions.

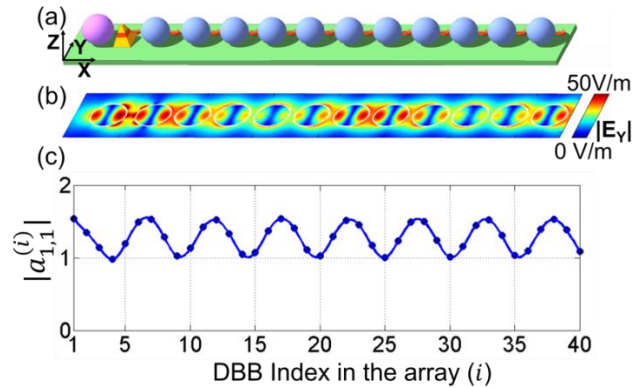

Fig.10. Integrated MTSQD-Nanoantenna-Waveguide Structure. (a) Schematic showing the integrated unit structure of the MTSQD interspersed with the nanoantenna-waveguide structure with spherical (GaAs) DBBs. The waveguide consists of array of 40 DBBs (blue) of radius 129 nm with center-center spacing of 300.6 nm. The reflector DBB (purple sphere) of the Yagi-Uda type nanoantenna geometry is of radius 131.7 nm. The QD is placed on the axis of the array, at 40 nm distance from the surface of both the blue and the purple DBBs. The transition dipole of the QD is chosen to be along Y direction. (b) The distribution of $|E_Y|$ component (normalized to 1 Debye strength of the QD transition dipole) of the electric field on the XY plane passing through the center of the DBBs, (c)The absolute value of the magnetic dipole mode coefficient ($|a_{1,1}|$) at 980 nm along the complete array of 40 DBBs, showing lossless propagation via the collective magnetic dipole mode. The oscillation of $|a_{1,1}|$ along the array arises from the Fabry-Perot interference.

The optical response of the assembly of the DBBs is analyzed employing Mie theory based multipole expansion method[69,70]. For certain symmetric DBBs (spherical, cylindrical) the multipole expansion method provides physical insight into the interaction between different multipole resonances in such systems. In this formulation, the scattered electromagnetic field is expressed in the basis of the magnetic and electric multipole modes for an arbitrary assembly of *N* DBBs of symmetric shape. For spherical shape of the DBBs, the multipole expansion of the electric field takes the following form[70],

$$\bar{E}(\bar{r}) = \sum_{i=1}^{N} \sum_{n=1}^{n_{max}} \sum_{m=-n}^{n} \left( a_{n,m}^{(i)} \bar{E}_{TE_{n,m}}(\bar{r} - \bar{r}_i) + b_{n,m}^{(i)} \bar{E}_{TM_{n,m}}(\bar{r} - \bar{r}_i) \right) \quad (4)$$

where $\{\bar{r}_i, i=1:N\}$ represent the spatial locations of the DBBs. $TE_{n,m}$ and $TM_{n,m}$ represents the magnetic multipole and electric multipole resonance modes of the individual DBBs, with the expansion coefficients in Equ. (4) denoted as $\left\{ a_{n,m}^{(i)}, b_{n,m}^{(i)} \mid i = 1:N; n = 1:n_{max}; m = -n:n \right\}$. As the indices n and m refer to the radial order (radial field distribution) and angular momentum (angular field distribution)[70] of the multipole modes, this approach allows truncating the multipole mode sequence in a controlled way depending on specific physical situations. In our case, the multipole expansion is truncated with $n_{max} = 4$ that limits the relative error in the calculated field values to ~$10^{-7}$ at 980nm. The frequency domain solution for the multipole mode coefficients for a specific source is obtained by analytically defining the multipole mode susceptibilities as well as mode-mode interaction matrices[70] followed by solving the acquired matrix equation. This formulation enables exploration of the collective resonances of different multipole modes of different symmetries and spectral responses that in turn allows design and optimization of multiple functions at multiple wavelengths in the DBB based LMU[29,30]. In the specific example shown in Fig.10(a), as noted above, the guiding and lossless propagation of the MTSQD emission at 980 nm is implemented by the collective excitation of the transverse magnetic dipole mode ($TE_{1,\pm1}$) of the blue DBBs to be at 980 nm. The distribution of the |E$_Y$| component of the electric field for this collective magnetic dipole excitation (normalized to 1 Debye strength of the QD transition dipole) on the XY plane passing through the center of the DBBs is shown in Figure 10(b). The lossless nature of the photon propagation in this collective mode is apparent from the plot of the absolute value of the $TE_{1,1}$ mode coefficient ($|a_{1,1}|$) along the complete array of 40 DBBs as shown in Figure 10 (c), where, apart from periodic undulations arising from the Fabry-Perot interference, the mode coefficients remain unchanged over the whole array.

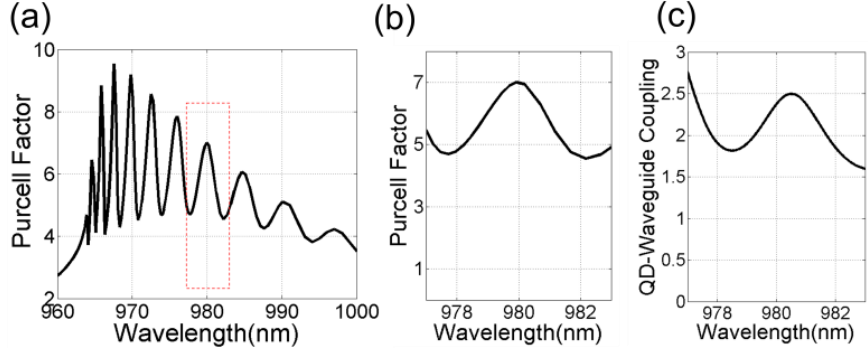

Figure 11. (a) Wavelength dependence of Purcell enhancement by the dielectric nanoantenna-waveguide structure over the complete propagation band. The non-uniformly spaced Fabry-Perot fringes originate from the dispersive nature of the propagation band. (b) The Purcell enhancement peak (inside the dotted rectangle in (a)) around the MTSQD emission wavelength at 980 nm, showing a factor of 7 enhancement with ~4 nm width. (c) Wavelength dependence of the MTSQD-waveguide coupling ($\Gamma_{MTSQD-WG}$) around the MTSQD emission wavelength. The coupling factor is defined in Eqn.(5).

The nanoantenna function of the LMU is implemented by the purple DBB acting as a reflector and the blue DBBs acting as directors in the well-known Yagi-Uda architecture[27]. The enhancement of the MTSQD emission rate (Purcell enhancement) produced by this nanoantenna structure is estimated by calculating the enhancement of the photon local density of states (LDOS) at the location of the QD employing the Green function based approach[71]. The Purcell enhancement as a function of wavelength is plotted in Figure 11(a) and shows Fabry-Perot fringes, signifying participation of the guiding modes in the enhancement of the photon LDOS. Here our LMU (i.e. the nanoantenna-waveguiding composite structure) is chosen in such a way that the desired operating wavelength (980 nm) lies sufficiently far from the slow edge of the propagating modes at 965 nm resulting in Purcell enhancement at 980 nm with a broad (~4 nm) spectral width (Figure 11(b)), suited for the spectral uniformity attainable via the SESRE approach. With alternative architectures, by specifically exploiting the highly dispersive slow guiding modes near the edge, Purcell enhancement can be boosted to ~hundred[28]. This effect is similar to the slow band-edge modes that have been exploited to demonstrate high Purcell factor in photonic crystal waveguide structures[1]. Thus with the enhanced Purcell effect in DBB based structures, reduction of the QD exciton lifetime from the typical ~1 ns to ~10 ps resulting in single photon emission up to 100 GHz repetition rate may be possible.

The emitted single photon is directed towards the DBB chain by the nanoantenna effect and coupled to the waveguide modes. The efficiency of the QD-waveguide coupling ($\Gamma_{MTSQD-WG}$) is expressed as the normalized magnetic dipole excitation of the terminal DBB of the waveguide with respect to the source electric dipole representing the transition in the MTSQD (expansion coefficient $b_{1,0}^{(source)}$), i.e.

$$\Gamma_{MTSQD-WG} = \frac{\left|a_{1,1}^{(N)}\right|^2 + \left|a_{1,-1}^{(N)}\right|^2}{\left|b_{1,0}^{(source)}\right|^2} \tag{5}$$

Figure 11(c) shows the wavelength dependence of $\Gamma_{MTSQD-WG}$ near the QD emission wavelength, demonstrating efficient on-chip energy transfer from the source dipole via the nanoantenna-waveguide LMU. Strong directionality is indicated by the higher than unity value of the efficiency $\Gamma_{MTSQD-WG}$, which arises from the Fabry-Perot resonances in the finite waveguide. Such nanoantenna-waveguide LMU provides on-chip simultaneous multifunctional control of SQD emission rate, propagation direction and efficiency of the emitted photons. The DBB based on-chip LMUs provide an alternative approach for building on-chip interconnected photonic networks with subwavelength size building blocks, thus smaller footprint compared to 2D photonic crystals. Experimental fabrication and characterization of DBB based LMUs and SQD-DBB integrated structure is underway.

## IV. Conclusion

The work reported here is a contribution to addressing the challenging goal of realizing on-chip integrated nanophotonic systems comprising spatially-regular array of single photon sources, each embedded in light manipulating elements (cavity, waveguide, beam-splitter, etc.), as units suited for creating optical circuits operational down to single photon level. Our efforts, and the results reported here, are guided by the continuing need for improved spatially ordered and spectrally uniform single photon emitter arrays that satisfy two additional requirements: (1) on-chip integrability with the LMU (comprising cavity, waveguide, beam splitter, etc.) and (2) the on-chip interconnectivity of such SQD-LMU units to, in turn, enable on-chip realization of functional optical circuits. To this end, the paper demonstrates the realization of a spatially ordered array of single quantum dot based single photon emitters with significantly improved (over current SAQDs) spectral uniformity created using the SESRE approach that is inherently on-chip integrable with subsequent fabrication of the above noted optical elements. Indeed the light manipulation elements can be created utilizing any of the current platforms- photonic crystals, micropillars, and plasmonics. Towards integration, the paper extends the aforementioned approaches and reports simulations of the optical response of DBB based multifunctional nanoantenna-waveguide that embed a QD and act as a LMU suitable for building on-chip optical circuits. Together, the two findings reported here provide incentive for further experimental and theoretical exploration and refinement of the approach.

Specifically, utilizing the substrate-encoded size-reducing epitaxy approach we have demonstrated an array of spatially-ordered GaAs/InGaAs single quantum dot based single photon sources emitting at 8K with a $g^{(2)}(0)$ of 0.19±0.03 and continuing to exhibit clear signature of single photon emission up to 77 K. The $g^{(2)}(0)$ values are limited by the resolution of the instrument and thus reflect upper limits. The high temperature emission for the InGaAs system reflects the high degree of three dimensional quantum confinement realized through control on growth. The created array of flat-top pyramidal shape MTSQDs with {103} side walls shows a spectral emission uniformity of 8.3 nm, nearly an order of magnitude tighter compared to SAQDs or NCQDs[4,63]. We note that no attempt has been made to optimize the QD growth in this first exploration and with further work such as the use of binary material[24] for QDs along with the full power of the already existing nanometer scale control on mesa size patterning and etching we expect the uniformity can be further improved significantly to be compatible with most of the external tuning approaches (Stark effect, piezoelectric tuning, etc.).

The shallow pyramidal SPS array created using SESRE can be easily planarized via appropriate overgrowth. Thus SESRE based MTSQDs lend themselves naturally to integration with light manipulating elements that can be fabricated subsequently using well-developed lithographic processes. To guide such integration we undertook simulation of on-chip integrated MTSQD with multifunctional (focusing, nanoantenna and waveguide) LMU (Fig.2 and Fig.10). The simulation is based upon the Mie scattering from the collective response of the multipoles (dominantly magnetic) of high refractive index subwavelength size dielectric building blocks. Utilizing GaAs spherical nanoparticle building blocks (the shape enables analytical modeling of optical response of multifunctional structures), we find, for 980 nm emitting SPS, a factor of ~7 enhancement in the MTSQD spontaneous emission rate, enhanced coupling to guiding modes, and lossless propagation of emitted photon. The SESRE approach to creating SPS arrays in a wide variety of material systems thus opens a powerful and rich path for realizing on-chip integrated quantum information processing architectures. Finally, we note that while these initial simulations have been carried out without accounting for the presence of the underlying substrate (as it enables analytical description), simulations including the substrate can be carried out employing FDTD approach. However, the well recognized fact that a substrate inherently induces adverse departures in the desired effects, real systems such as integrated SQD-photonic crystal based cavity / waveguide, are already being examined in a suspended membrane configuration[1,19] and the same will be true for the dielectric building block approach proposed here to be explored.

## ACKNOWLEDGMENTS

This work is supported by ARO grant number W911NF-15-1-0298 (Program Manager: Dr. John Prater). Work on the growth of the quantum dots was supported by AFOSR grant number FA9550-10-01-0066.